\documentstyle[twoside,fleqn,espcrc2,epsf]{article}
\newcommand{\AmS}{{\protect\the\textfont2
  A\kern-.1667em\lower.5ex\hbox{M}\kern-.125emS}}

\def\bfnabla{\mbox{\boldmath $\nabla$}}

\def\bfsigma{\mbox{\boldmath $\sigma$}}
\def\lQ{\Lambda_{QCD}}
\newcommand{\nn}{\nonumber}
\newcommand{\be}{\begin{equation}}
\newcommand{\ee}{\end{equation}}
\newcommand{\bea}{\begin{eqnarray}}
\newcommand{\eea}{\end{eqnarray}}

\def\als{\alpha_{s}}
\def\siml{{\
    \lower-1.2pt\vbox{\hbox{\rlap{$<$}\lower6pt\vbox{\hbox{$\sim$}}}}\ }} 

\title{Heavy Quarkonium Potential and inclusive decay widths in terms
of Wilson loops} 

\author{Antonio Pineda\address{Dept. d'Estructura i Constituents de la
  Mat\`eria and IFAE, U. Barcelona \\ Diagonal 647, E-08028 Barcelona,
  Catalonia, Spain} }

\begin{document}
 
\begin{abstract}
We briefly summarize some results obtained by potential NRQCD in the
non perturbative regime: 1) a systematic procedure to obtain the heavy
quarkonium potential in the $1/m$ expansion (with explicit expressions
up to $O(1/m^2)$ in terms of Wilson loops), and 2) a description of
the heavy quarkonium inclusive decays to light particles, and of the
NRQCD matrix elements (also the octet ones), in terms of the
wave-function at the origen and some gluonic field-strength
correlators.
\end{abstract}

\maketitle

Heavy quarkonium ($b$-$\bar b$, $c$-$\bar c$) systems have been
traditionally described by potential models in the past, being their
inverse size assumed to be of $O(\lQ)$ (and that $\lQ \ll m$, being
$m$ the mass of the heavy quark).  Potential models are characterized
by the introduction of a, more or less, phenomenological potential in
a Schr\"odinger equation. By assuming some functionality in $r$ and by
fitting the free parameters of the potential, a good description of
the heavy quarkonium spectrum was obtained. Nevertheless, there were
two issues: 1) under which circumstances, and how, a pure
Schr\"odinger formulation will emerge from QCD in the non-perturbative
regime and, if so, 2) how to obtain the potentials from QCD, or, at
least, how to relate them with objects eventually computable in QCD
(then any potential model should, at least, be consistent with points
1) and 2)).  The use of effective field theories has helped to clarify
when point 1) is satisfied and how it can be derived from QCD (see
\cite{pNRQCD,M1,M2}). The resulting Lagrangian of the effective theory
(that we name potential NRQCD (pNRQCD)) reads
\be
{\cal L}_{\rm pNRQCD} = S^\dagger 
\bigg( i\partial_0 -h({\bf x}_1,{\bf x}_2, {\bf p}_1, {\bf p}_2)\bigg) S, 
\label{pnrqcdl}
\ee
where $h$ is the Hamiltonian of the singlet, i.e. of the heavy
quarkonium.  Actually $h$ is only a function of ${\bf r}={\bf
x}_1-{\bf x}_2$, ${\bf p}_1$, ${\bf p}_2$, which is analytic in the
two last operators but contains non-analyticities in {\bf r}. ${\bf
p}_1= -i \bfnabla_{{\bf x}_1}$ and ${\bf p}_2= -i \bfnabla_{{\bf
x}_2}$. $h$ has the following expansion up to order $1/m^2$:
\bea
\label{hss}
&&
h ({\bf x}_1,{\bf x}_2, {\bf p}_1, {\bf p}_2) = 
{{\bf p}^2_1\over 2 m_1} +{{\bf p}^2_2\over 2 m_2} 
+ V^{(0)}
\\
\nn
&&
+{V^{(1,0)} \over m_1}+{V^{(0,1)} \over m_2}+ {V^{(2,0)} \over m_1^2}
+ {V^{(0,2)}\over m_2^2}+{V^{(1,1)} \over m_1m_2}.
\eea
For the $V^{(2,0)}$ and $V^{(0,2)}$ potentials we define
\be
V^{(2,0)}=V^{(2,0)}_{SD}+V^{(2,0)}_{SI}, \;
V^{(0,2)}=V^{(0,2)}_{SD}+V^{(0,2)}_{SI}. 
\ee
The spin-independent terms can be written as 
\be
V^{(2,0)}_{SI}=\left\{{{\bf p}_1^2 \over 2},V_{{\bf p}^2}^{(2,0)}(r)\right\}
+{V_{{\bf L}^2}^{(2,0)}(r)\over r^2}{\bf L}_1^2 + V_r^{(2,0)}(r),
\ee
and
\be
V^{(0,2)}_{SI}=\left\{{{\bf p}_2^2 \over 2},V_{{\bf p}^2}^{(0,2)}(r)\right\}
+{V_{{\bf L}^2}^{(0,2)}(r)\over r^2}{\bf L}_2^2 + V_r^{(0,2)}(r),
\ee
where ${\bf L}_1 \equiv {\bf r} \times {\bf p}_1$ and ${\bf L}_2 
\equiv {\bf r} \times {\bf p}_2$ and
\bea
\nn
&&
V_{{\bf p}^2}^{(2,0)}(r) =V_{{\bf p}^2}^{(0,2)}(r), \; 
V_{{\bf L}^2}^{(2,0)}(r) =V_{{\bf L}^2}^{(0,2)}(r), 
\\
&&
V_r^{(2,0)}(r)=V_r^{(0,2)}(r;m_2 \leftrightarrow m_1). 
\eea
The spin-dependent part of $V^{(2,0)}$ is of the type 
\be
V^{(2,0)}_{SD}=V^{(2,0)}_{LS}(r){\bf L}_1\cdot{\bf S}_1.
\ee
Analogously, for the $V^{(0,2)}$ potential we can write 
\be
V^{(0,2)}_{SD}=-V^{(0,2)}_{LS}(r){\bf L}_2\cdot{\bf S}_2, 
\ee
where  $V^{(2,0)}_{LS}(r)=V^{(0,2)}_{LS}(r; m_2 \leftrightarrow m_1)$.\\
For the $V^{(1,1)}$ potential we define
\be
V^{(1,1)}=V^{(1,1)}_{SD}+V^{(1,1)}_{SI}.
\ee
The spin-independent part can be written as 
\bea
&&
V^{(1,1)}_{SI}= -{1 \over 2}\left\{{\bf p}_1\cdot {\bf p}_2,V_{{\bf
p}^2}^{(1,1)}(r)\right\} 
\\
\nn
&&
-{V_{{\bf L}^2}^{(1,1)}(r)\over 2r^2}({\bf L}_1\cdot{\bf L}_2+ {\bf
L}_2\cdot{\bf L}_1)+ V_r^{(1,1)}(r), 
\eea
while the spin-dependent part contains the following operators:
\bea
\nn
&&
V^{(1,1)}_{SD}=
V_{L_1S_2}^{(1,1)}(r){\bf L}_1\cdot{\bf S}_2 -
V_{L_2S_1}^{(1,1)}(r){\bf L}_2\cdot{\bf S}_1 
\\
&&
+ V_{S^2}^{(1,1)}(r){\bf S}_1\cdot{\bf S}_2 + V_{{\bf
S}_{12}}^{(1,1)}(r){\bf S}_{12}({\hat {\bf r}}),  
\eea
where ${\bf S}_{12}({\hat {\bf r}}) \equiv 3 {\hat {\bf r}}\cdot
\bfsigma_1 \,{\hat {\bf r}}\cdot \bfsigma_2 - \bfsigma_1\cdot \bfsigma_2$ and 
$V_{L_1S_2}^{(1,1)}(r)=V_{L_2S_1}^{(1,1)}(r; m_1 \leftrightarrow m_2)$. 

\medskip

Point 2), i.e. how to obtain the relation of the potential with
objects computable within QCD, can be summarized in the following
question:

\medskip

{\it How to obtain the potential in terms of Wilson loops in the $1/m$
expansion (also usually named adiabatic or Born-Oppenheimer
approximation)?}

\medskip

The first attempts to answer this question started more than twenty
years ago. The expression for the leading spin-independent potential,
of $O(1/m^0)$, corresponds to the static Wilson loop and was derived
by Wilson and Susskind \cite{WS}:
\be 
V^{(0)}(r) = \lim_{T\to\infty}{i\over T} \ln \langle W_\Box \rangle.  
\label{v0}
\ee
Expressions for the leading spin-dependent potentials in the $1/m$
expansion, of $O(1/m^2)$, were given in Refs. \cite{spin}.  The
procedure followed in these works proved to be very difficult to
extend beyond these leading-order potentials. In Ref. \cite{BMP}, a
new method to calculate the potentials was proposed, where new
spin-independent (some of them momentum-dependent) potentials at
$O(1/m^2)$ were obtained. In \cite{states}, expressions for the
spin-dependent potentials were obtained in terms of eigenstates of the
static limit of the NRQCD Hamiltonian in the Coulomb gauge. In these
works, the potentials did not correctly reproduce the
ultraviolet behaviour expected from perturbative QCD (the hard logs
$\sim \log m$). This was the first signal that a controlled derivation
of the potentials from QCD was needed. The solution to this problem
needs of NRQCD \cite{nrqcd} where the ultraviolet behavior is encoded
in the matching coefficients of the NRQCD operators. It is then
possible to incorporate them to the potentials as done in
\cite{coeffNRQCD}. At that point, the obtained set of potentials at
$O(1/m^2)$ seemed to be complete.

Nevertheless, this view has recently been challenged in
Refs. \cite{M1,M2}, where a systematic study of the potential has
been done within an effective field theory framework: pNRQCD.  The
main improvements achieved in Refs. \cite{M1,M2} with respect these
previous computations can be summarized as follows:
\begin{itemize}
\item[A)]We have developed a general procedure to compute the
potential by equating green functions in NRQCD and pNRQCD \cite{M1}.
\item[B)]We have developed the general method and formally provide
recursive equations to obtain the potential at any order in $1/m$ in
terms of matrix elements and energies of the states solution of the
static limit \cite{M1,M2}. These expressions can then be rewritten in
terms of Wilson loops.
\end{itemize}

Points A) and B) solve, in alternative ways, question 2) and,
thus, finally settle this issue, opened since more than twenty years
ago.

Once the formalism has been developed, we have been able to obtain,
for the first time, the {\it complete} potential (up to field
redefinitions) in pure gluedynamics up to $O(1/m)$ in \cite{M1} and up
to $O(1/m^2)$ in \cite{M2}. If compared with the previous expressions
for the potential that one may find in the literature, our novel
findings can be summarized as follows:
\begin{itemize}
\item[C)] A new  $O(1/m)$ potential \cite{M1}.
\item[D)] Correct expressions for the spin/velocity-independent terms of
the $1/m^2$ potential: $V_r^{(0,2)}$ ($V_r^{(2,0)}$) and $V_r^{(1,1)}$
\cite{M2}. 
\item[E)] We have clarified that the, so far, {\it standard} expression
obtained by Eichten and Feinberg for $V_{L_1S_2}^{(1,1)}$
($V_{L_2S_1}^{(1,1)}$) and later {\it confirmed} by
several groups was incorrect \cite{M2}.
\end{itemize}

Let us stress that, to date, A) and B) are the only available methods
in the literature to compute the potential in terms of Wilson loops
within a {\it systematic} expansion in $1/m$. The attempts to
implement the method of Eichten and Feinberg beyond their
leading-order results were not able to obtain finite expressions
\cite{thesis}. Indeed, in a way, the procedure A) can be seen as the
generalization of the Eichten and Feinberg method. In order to obtain
this generalization it was crucial to understand the computation
within an effective field theory ideology where equalities between
Green functions were imposed and interpolating fields with arbitrary
normalizations used. The method advocated in Ref. \cite{BMP} does not
appear to be correct, at least in its current formulation, since, for
instance, it is not able to obtain the $1/m$ potential. The
computations in Ref. \cite{states} essentially provide the correct
expressions for the spin-dependent potentials (once one takes the
NRQCD matching coefficients to tree level and neglects the tree-level
annihilation contribution in the equal mass case). Nevertheless, their
methodology needs to be generalized (along the lines of \cite{M1,M2})
to take into account the fact that one is dealing with operators
instead that with numbers in these type of computations.

Let us also notice that the expressions for the potential in terms of
Wilson loops encode the full perturbative series in $\als$, at leading
order in the multipole expansion, in the perturbative situation, which
is a well defined limit. 

Once $h$ has been obtained, we can obtain the energies of
the bound states. At the order of interest, we take the energies from
the real part of the Schr\"odinger equation 
\be
\label{Schr}
({\rm Re}\, h)\,\langle {\bf r}|n,l,s,j \rangle=E_{njls}\,\langle {\bf
r}|n,l,s,j \rangle, 
\ee 
with quantum numbers $n$, $j$, $l$ and $s$ as defined in
Ref. \cite{swave}.  From the imaginary piece of $h$, we can obtain the
inclusive decay widths (to light hadrons, leptons or
photons) by using the relation
\be
\Gamma= - 2 \, \langle n,l,s,j| {\rm Im}\, h  |n,l,s,j \rangle.
\label{imag}
\ee
This is what we have done in Refs. \cite{pwave,swave} on which we
report in what follows.

The imaginary parts of $h$ are inherited from the four-fermion
matching coefficients of NRQCD: ${\rm Im}\,c_{4-f}^{d=6}$, ${\rm
Im}\,c_{4-f}^{d=8}$, .... Therefore, the structure of the imaginary
part of the Hamiltonian reads\\
$
\displaystyle{
{\rm Im}\,h \sim 
\sum_k{\rm Im}\,c_{4-f,k}^{d=6}{\delta^{(3)}({\bf r}) \over m^2}
\left(A_k+B_k{\lQ^2 \over m^2}+\cdots \right)
}$
\smallskip
$
\displaystyle{
+
\sum_k{\rm Im}\,c_{4-f,k}^{d=8}\left(A'_k{\{\delta^{(3)}({\bf
r}),\bfnabla^2\}\over m^4} 
\right.
}$
\smallskip
$
\displaystyle{
\qquad
+
\left.
B_k'{ \delta^{(3)}({\bf x})\over m^2}{\lQ^2 \over m^2}+\dots \right)
}$
\smallskip
$
\displaystyle{
+
\sum_k{\rm Im}\,c_{4-f,k}^{d=8}
\left(
C_k{\mathcal{T}}_{SJ}^{ij}\frac{{\bfnabla}^i 
\delta^{(3)}({\bf r}) {\bfnabla}^j}{m^4}+\dots \right)
+ \dots \,,
}$ 
being dictated by local potentials. The S-wave and P-wave inclusive 
decay width easily follow 
\bea
\label{gammas}
&&
\Gamma_{S-{\rm wave}}\sim 
\sum_k{\rm Im}\,c_{4-f,k}^{d=6}{|R_{ns0s}(0)|^2 \over m^2}
\\
&&
\quad\qquad
\times
\left(A_k+B_k{\lQ^2 \over m^2}+\cdots \right)
\nn
\\
\nn
&&
\quad
+
\sum_k{\rm Im}\,c_{4-f,k}^{d=8}\left(A'_k{R_{ns0s}(0)(\bfnabla^2
R_{n10s}(0))\over m^4} 
\right.
\\
\nn
&&
\quad\qquad
\left.
+B'_k{|R_{ns0s}(0)|^2 \over m^2}{\lQ^2 \over m^2}+\dots \right)
+
\dots \,,
\eea
\be
\label{gammap}
\Gamma_{P-{\rm wave}}
\sim
\sum_k{\rm Im}\,c_{4-f,k}^{d=8}
C_k{|\bfnabla R_{nj1s}(0)|^2 \over m^4} +
\dots , 
\ee
where $R_{ns0s}$ and $R_{nj1s}$ are the $S$-wave and $P$-wave radial
component of $\langle {\bf r}|n,l,s,j \rangle$.

Equations. (\ref{gammas}-\ref{gammap}) illustrate the factorization of the
soft and ultrasoft scale since the coefficients $A_k$, $B_k\lQ^2$,
$C_k$, etc. are independent of the bound-state dynamics.  This makes
possible to obtain model independent predictions from the above
results, which just come out from the structure of our effective
theory and the fact that the imaginary pieces are encoded in the
four-fermion matching coefficients. In any case we are able to go
further and to obtain the coefficients $A_k$, $B_k\lQ^2$, $C_k$,
etc. in terms of gluonic correlators or the identity (a specific case
of gluonic correlator).  Therefore we can relate them with QCD. These
correlators could be eventually computed in the lattice, evaluated
with models or fixed with the experiment.  For illustration, we show
the non-perturbative parameter which appears in $P$-wave decays
\cite{pwave}:
\newpage
\be
C_k \sim {1 \over N_c}\int_0^\infty dt \, t^3 \langle g{\bf E}(t)\cdot
g{\bf E}(0)\rangle 
\,.
\ee
The explicit expressions for the gluonic correlators can be found in
Refs. \cite{pwave,swave}, where Eqs. (\ref{gammas}-\ref{gammap}) have
been obtained with relative precision $O(\lQ^2/m^2)$.

The traditional description of the inclusive decays of heavy
quarkonium within an effective field theory framework is in terms of
expectation values of the 4-fermion operators of NRQCD
\cite{nrqcd}. At this respect our procedure can give a master equation
which relates these matrix elements with wave-functions at the origen
and some non-perturbative constants independent of the mass and the
bound state dynamics. Given a four-fermion operator $O$, the master
equation is the following ($|H\rangle$ represents a generic
heavy-quarkonium state at rest, ${\bf P}=0$):
\bea
\nn
&&\langle H|O({\bf 0})|H\rangle ={1\over \langle {\bf P}=0| {\bf P}=0 \rangle }
\int d^3{\bf r}\int d^3{\bf r}' 
\\
\nn
&&
\times
\int d^3{\bf R}\int d^3{\bf R}'
\langle {\bf P}=0|{\bf R}\rangle \langle njls |{\bf r}\rangle
\\
\nn
&&
\times
\bigg[
\langle \underbar{0}; {\bf x}_1 {\bf x}_2|
\int d^3\xi O({\bf \xi}) \;
| \underbar{0}; {\bf x}_1^\prime {\bf x}_2^\prime \rangle
\bigg]
\\
&&
\times
\langle {\bf R}'|{\bf P}=0\rangle \langle {\bf r}'|njls \rangle
\,,
\eea
where ${\bf r} = {\bf x}_1-{\bf x}_2$, ${\bf r}' = {\bf x}'_1-{\bf
x}'_2$, ${\bf R} = ({\bf x}_1+{\bf x}_2)/2$ and ${\bf R}' = ({\bf
x}'_1+{\bf x}'_2)/2$ (note that $\langle {\bf R}'|{\bf P}=0\rangle=1$
and $\langle {\bf P}=0| {\bf P}=0 \rangle=\int d^3x$). The state $|
\underbar{0}; {\bf x}_1 {\bf x}_2 \rangle$ represents the heavy
quark-antiquark ground state of the QCD Hamiltonian {\it in the $1/m$
expansion} (see Ref. \cite{swave} for details).
 
It is especially important that we can describe octet matrix elements,
which were thought that could not be related with a Schr\"odinger-like
formulation in any way. We have shown that this is not so.
Nevertheless, we would like to stress that even for the singlet matrix
elements, which were thought to be related with the wave-function at
the origen of some Schr\"odinger-like formulation, our results are not
trivial because no such formulation existed in the non-perturbative
case before. We have provided it, formalizing the connection between
NRQCD matrix elements and wave-functions computed in a Schr\"odinger
equation and showing how this connection can be improved in a
systematic way.
 
Our results allow to reduce significantly, by a factor of two, the
number of non-perturbative constants compared with the number of NRQCD
matrix elements if we assume that all the states below the $D$-$\bar
D$ and $B$-$\bar B$ production threshold are in the dynamical
situation $\lQ \gg mv^2$. In particular, one can think of combinations
that are free of non-perturbative effects (see \cite{swave}).

I would like to thank N. Brambilla, D. Eiras, J. Soto and A. Vairo for
collaboration on this topic.

\end{document}